\documentclass[letter]{IEEEtran}

\usepackage{graphicx,color}
\usepackage{cite}
\usepackage{setspace} 
\usepackage{amsmath,amsthm,amssymb}
\usepackage{multirow}
\usepackage{hhline}
\usepackage{epsfig}
\usepackage{epstopdf}
\usepackage{verbatim}
\usepackage{algorithm}
\usepackage{algpseudocode}
\usepackage{cases}
\usepackage{array}
\usepackage{subfigure}
\allowdisplaybreaks

\usepackage{forloop}
\newcounter{ct}

\makeatletter 
\def\@eqnnum{{\normalsize \normalcolor (\theequation)}} 
\makeatother
\begin{document}

\title{Joint Optimal Design for Outage Minimization in DF Relay-assisted {Underwater Acoustic Networks}}
\author{Ganesh Prasad, \textit{Member, IEEE}, Deepak Mishra, \textit{Member, IEEE}, and Ashraf Hossain, \textit{Senior Member, IEEE}
    \thanks{G. Prasad and A. Hossain are with the Department of Electronics and Communication Engineering, National Institute of Technology, Silchar, India (e-mail: \{gpkeshri, ashraf\}@ece.nits.ac.in).
        
    D. Mishra is with the Department of Electrical Engineering, Link\"oping University, Link\"oping 58183, Sweden (e-mail: deepak.mishra@liu.se).
    }}

\maketitle
\begin{abstract}
	 This letter minimizes outage probability in a single decode-and-forward (DF) relay-assisted underwater acoustic network (UAN) without direct  source-to-destination link availability. \color{black} Specifically, a joint global-optimal design for relay positioning and allocating power to source and relay is proposed. For analytical insights, a novel low-complexity tight approximation method is also presented. Selected numerical results validate the analysis and quantify the comparative gains achieved using optimal power allocation (PA) and relay placement (RP) strategies.   
\end{abstract}   


\section{Introduction}
\label{sec_introduction} 
 Due to their prominent applications, the underwater acoustic networks (UANs) have gained significant research interest~\cite{dar01}. However, the data rate in UANs is limited due to eminent delay and restricted bandwidth over long range communications. Therefore, if  source to destination direct link is incompetent to meet a data rate demand, then a relay can be deployed between them to decrease the hop length and yield an energy efficient design~\cite{stoj02}.  This letter investigates the joint power allocation (PA) and relay placement (RP) in a dual-hop UAN where the direct link is either absent~\cite{wan13}, or its effect can be neglected while minimizing the outage probability for a desired date rate.\color{black}
	
In the recent works~\cite{liu04} and~\cite{kam09}, an energy efficient UAN operation was investigated by optimizing the location of the relays along with other key parameters. Whereas, optimal PA  was studied in~\cite{babu10}. Although multiple relays were used in these works, the underlying optimization studies were performed considering assumptions like perfect channel state information (CSI) availability and adopting simpler Rayleigh fading model. In contrast, the joint optimization in this letter has been carried out under a realistic dual-hop communication environment~\cite{nour11}, where only the statistics of fading channels are required and a more generic Rician distribution is adopted for the frequency-selective fading channel. Lately, in~\cite{nour11,cao12,doos07} it is shown that the throughput in cooperative UANs can be significantly improved by optimizing PA and RP. However, the existing works didn't consider joint optimization and also only numerical solutions were proposed for individual PA and RP problems. \textit{So, to the best of our knowledge, the joint global optimization of PA and RP in UANs has not been investigated}. Further, we would like to mention that the joint optimization in cooperative UANs is very different and more challenging than the conventional terrestrial networks due to the frequency-selective behavior of underwater channels in terms of fading, path loss and noise, which are all strongly influenced by the operating frequencies.

The key contributions of this letter are three fold. First we prove the generalized convexity of the proposed outage minimization problem in DF relay-assisted UANs. Using it we obtain the jointly global optimal PA and RP solutions. Secondly, to gain analytical insights, a novel very low-complexity near-optimal approximation algorithm is presented. Lastly via numerical investigation, the analytical discourse is first validated and then used for obtaining insights on the optimizations along with the quantification of achievable performance gains.
 
\section{System Model Description}\label{section2}
We consider a dual-hop, half-duplex DF relay assisted UAN. Here  a source $\mathcal{S}$ communicates with destination $\mathcal{D}$, positioned at $D$ distance apart, via a cooperative relay $\mathcal{R}$. These nodes are composed of single antenna and the $\mathcal{S}$-to-$\mathcal{D}$ direct link is not available due to large path loss and fading effects.  
As $\mathcal{R}$ communicates in half-duplex mode, the data transfer from $\mathcal{S}$ to $\mathcal{D}$ takes place in two slots: first from $\mathcal{S}$ to $\mathcal{R}$ and then from $\mathcal{R}$ to $\mathcal{D}$.  For efficient energy utilization, a power budget $P_B$ is taken for transmit powers of $\mathcal{S}$ and $\mathcal{R}$. We assume that each of $\mathcal{S} \mathcal{R}$, $\mathcal{R} \mathcal{D}$, and $\mathcal{S} \mathcal{D}$ links follows independent Rician fading.
      
Adopting the channel model in~\cite{cao12,stoj02}, the frequency $f$ dependent received signal-to-noise ratio (SNR) at node $j$, placed $d_{ij}$ distance apart from node $i$, is given by: 
\begin{equation} 
		\gamma_{ij}(f)=S_{i}(f)G_{ij}(f)\left[a(f)\right]^{-d_{ij}}d_{ij}^{-\alpha}\left[N(f)\right]^{-1}.
\end{equation}
Here $G_{ij}(f)$ is the channel gain for frequency-selective Rician fading over $ij$ link,  $S_i(f)$ is power spectral density (PSD) of transmitted signal from node $i$, $\alpha$ is  spreading factor, $a(f)$ is  absorption coefficient in dB/km for $f$ in kHz~\cite[eq. (3)]{stoj02}, and $N(f)$ is PSD of noise as defined by~\cite[eq. (7)]{stoj02}. \color{black} The complementary cumulative distribution function (CDF) of $\gamma_{ij}$ for Rice factor $K \!\leq\!39$dB is approximated as~\cite[eq. (10)]{mish17}: 
\begin{equation}\label{ccdf}
	\begin{aligned}
		\textbf{Pr}[\gamma_{ij}(f)>x]=e^{-\mathcal{A}\Big(\frac{2(1+K)\beta x}{\overline{\gamma}_{ij}(f)}\Big)^\mathcal{B}},
	\end{aligned}
\end{equation}
where $\mathcal{A}=e^{\phi(\sqrt{2K})}$ and $\mathcal{B}=\frac{\varphi(\sqrt{2K})}{2}$. The polynomial expressions for $\phi(v)$ and $\varphi(v)$ as a function of $v$ were defined in~\cite[eqs. (8a), (8b)]{mish17}. The expectation of SNR $\gamma_{ij}(f)$ is given by  $\overline{\gamma}_{ij}(f)=\frac{\beta c_{ij}(f)S_{i}(f)}{N(f)a(f)^{d_{ij}}d_{ij}^{\alpha}}$, where $\beta=\frac{\mathcal{A}^{-\frac{1}{\mathcal{B}}}}{\mathcal{B}}\Gamma(\frac{1}{\mathcal{B}})$ and $c_{ij}(f)$ is the expectation of the channel gain $G_{ij}(f)$. Using this channel distribution information (CDI), we aim to minimize the outage probability for $\mathcal{S} \mathcal{D}$ underwater communication. 
 
\section{Joint Optimization Framework}\label{optimization} 
Here we first obtain the outage probability expression and then present the proposed joint global optimization framework.  

\subsection{Outage Minimization Problem}  
Outage probability  is defined as the probability of received signal strength falling below an outage data rate threshold $r$.

\noindent Outage probability $p_{out}$ in DF relay without direct link is~\cite{mish17}:
\begin{eqnarray}\label{out_exp} 
p_{out}=\textbf{Pr}\Big(\hspace{-1mm} \textstyle\int\limits_{0}^{B_W}\hspace{-1mm}\textstyle\frac{1}{2}\log_2(1+\min\{\gamma_{\mathcal{S}\mathcal{R}}(f),\gamma_{\mathcal{R}\mathcal{D}}(f)\})\text{d}f\leq r\hspace{-1mm}\Big) \hspace{-3mm} 
\end{eqnarray} 
	Our goal of minimizing $p_{out}$ by jointly optimizing PA and RP for a given transmit power budget can be formulated as below. 		
	\begin{eqnarray}\label{eq9}
	\begin{aligned}
	\text{(P0):}&\hspace{-0mm} \underset{S_\mathcal{S}(f),S_\mathcal{R}(f),d_{\mathcal{S}\mathcal{R}}}{\text{minimize}}
\quad p_{out}, \quad \text{subject to} \hspace{2mm}C1\hspace{0mm}: d_{\mathcal{S}\mathcal{R}} \geq \delta,\\
	& \hspace{-6mm} C2\!: d_{\mathcal{S}\mathcal{R}}\leq D-\delta, \hspace{1mm}C3\!: \textstyle\int_{0}^{B_W}\hspace{-0.5mm}(S_\mathcal{S}(f)+S_\mathcal{R}(f))\text{d}f\leq P_B,\hspace{-6mm}
	\end{aligned}
	\end{eqnarray} 
	where $C1$ and $C2$ are the boundary conditions on  $d_{\mathcal{S}\mathcal{R}}$ with $\delta$ being the minimum separation between two nodes~\cite{mish17}. $C3$ is the total transmit power budget in which $S_\mathcal{S}(f)$ and $S_\mathcal{R}(f)$ at frequency $f$ respectively represent the power spectral density (PSD) of transmit powers for $\mathcal{S}$ and $\mathcal{R}$. From the convexity of $C1,C2,C3$ along with the pseudoconvexity of $p_{out}$ in $S_\mathcal{S}(f)$, $S_\mathcal{R}(f)$, and $d_{\mathcal{S}\mathcal{R}}$ as proved in Appendix~\ref{AppA}, (P0) is a generalized-convex problem possessing the unique global optimality property~\cite[Theorem 4.3.8]{Baz06}. However, as it is difficult to solve $\text{(P0)}$ in current form, we next present an equivalent formulation to obtain the jointly optimal design. 
	 
\subsection{Equivalent Formulation for obtaining Joint Solution}\label{sec:Eqv}
As direct solution of (P0) is intractable~\cite{babu10,nour11}, we discretize the continuous frequency domain problem  (P0). For this transformation we choose the large enough number  $n$ of frequency sub-bands or ensure that the bandwidth of each sub-band $\Delta f=\frac{B_W}{n}$ is sufficiently small such that the difference between outage probabilities, $p_{out}$ defined in \eqref{out_exp} and $\widehat{p_{out}}$ defined in \eqref{eq:pouH} for the  discrete domain, have the corresponding root mean square error less than $0.08$ for it being a good fit~\cite{Hoop08}. So, instead of minimizing $p_{out}$, we minimize  
\begin{eqnarray}\label{eq:pouH}
\widehat{p_{out}}\triangleq\textstyle\textbf{Pr}\left( \sum_{q=1}^{n}\frac{\Delta f}{2}\log_2(1+\min\{\gamma_{\mathcal{S}\mathcal{R}_q},\gamma_{\mathcal{R}\mathcal{D}_q}\})\leq r\right)\hspace{-1mm}, \hspace{-3mm}
\end{eqnarray}
where $q^{th}$ sub-band of the $\mathcal{S}\mathcal{R}$ link is coupled with $q^{th}$ sub-band of the $\mathcal{R}\mathcal{D}$ link, the end-to-end received SNR at node  $j$ is: $\gamma_{{ij}_q}=P_{{i}_q}G_{{ij_q}}a_q^{-d_{ij}}d_{ij}^{-\alpha}[N_q\Delta f]^{-1}$, where $P_{{i}_q}=S_{{i}_q}\Delta f$ and $S_{{i}_q}=S_{i}(f_q)\text{U}(f-f_q)$ are the PA and PSD respectively at transmitting node $i\in\{\mathcal{S},\mathcal{R}\}$ with unit step function  $\text{U}(f)=1$ for $f\in [-\frac{\Delta f}{2},\frac{\Delta f}{2}]$ and $0$ otherwise. Further, $a_q=a(f_q)\text{U}(f-f_q)$ is the absorption coefficient, $N_q=N(f_q)\text{U}(f-f_q)$ is the additive noise, $G_{{ij}_q}=G_{ij}(f_q)\text{U}(f-f_q)$ and $c_{{ij}_q}=\mathbb{E}[G_{{ij}_q}]=c_{ij}(f_q)\text{U}(f-f_q)$ respectively are the channel gain and its expectation value in $q^{th}$ sub-band of ${ij}\in\{\mathcal{S}\mathcal{R},\mathcal{R}\mathcal{D}\}$ link. 
The different frequency-dependent parameters (cf. Section~\ref{section2}) remains constant within a sub-band and they are expressed by their respective center frequencies $\{f_q\}_{q=1}^n$. The twofold benefit of this discretization are transforming: (i) a frequency-selective fading channel into a non-frequency-selective one, and (ii) non-additive noise into an additive noise~\cite{babu10}. 

For sufficiently large value of $n$,  $\widehat{p_{out}}$ closely matches $p_{out}$ (as also shown later via Fig.~\ref{fig:validation}(a)), using Appendix~\ref{AppA}, we can claim that $\widehat{p_{out}}$ is also jointly-pseudoconvex in $\{P_{\mathcal{S}_q},P_{\mathcal{R}_q}\}_{q=1}^n$, and $d_{\mathcal{S}\mathcal{R}}$. Further, as CDF is a monotonically

\noindent decreasing function of the expectation of the underlying random variable~\cite[Theorem 1]{mish18} in \eqref{eq:pouH}, the minimization of $\widehat{p_{out}}$ is equivalent to the maximization of the expectation value $\frac{\Delta f}{2}\mathbb{E}[\log_2\prod_{q=1}^{n}(1+\min\{\gamma_{\mathcal{S}\mathcal{R}_q},\gamma_{\mathcal{R}\mathcal{D}_q}\})]$. Further, we observe that since the logarithmic transformation is monotonically increasing, expectation $\mathbb{E}[\prod_{q=1}^{n}(1+\min\{\gamma_{\mathcal{S}\mathcal{R}_q},\gamma_{\mathcal{R}\mathcal{D}_q}\})]$ is also a jointly pseudoconcave function. Lastly, assuming SNRs in different sub-bands to be independently and identically distributed, the products in this expectation can be moved outside the operator $\mathbb{E}\left[\cdot\right]$ and (P0) can be equivalently formulated as
\begin{equation}
\begin{aligned}
\text{(P1):}&\hspace{-0mm} \underset{\{P_{\mathcal{S}_q},P_{\mathcal{R}_q}\}_{q=1}^n,d_{\mathcal{S}\mathcal{R}}}{\text{maximize}}
&&\hspace{-2mm}  \textstyle{\prod\limits_{q=1}^{n}(1+\mathbb{E}[\min\{\gamma_{\mathcal{S}\mathcal{R}_q},\gamma_{\mathcal{R}\mathcal{D}_q}\}])} \\
&\hspace{-2mm} \text{subject to}
&&\hspace{-12mm} C1, C2, \widehat{C3}\hspace{-0.5mm}: \textstyle\sum_{q=1}^{n}(P_{\mathcal{S}_q}+P_{\mathcal{R}_q})\leq P_B,
\end{aligned}\hspace{-2mm}
\end{equation}     
where $\widehat{C3}$ gives the transmit power budget and using the definition~\eqref{meanSNR} in Appendix~\ref{AppA}, $\mathbb{E}[\min\{\gamma_{\mathcal{S}\mathcal{R}_q},\gamma_{\mathcal{R}\mathcal{D}_q}\}]=\overline{\gamma}_q\!=\!\textstyle\frac{\beta}{N_q}\Big[\Big(\frac{a_q^{d_{\mathcal{S}\mathcal{R}}}d_{\mathcal{S}\mathcal{R}}^\alpha}{c_{\mathcal{S}\mathcal{R}_q}P_{\mathcal{S}_q}}\Big)^\mathcal{B}+\Big(\frac{a_q^{D-\delta-d_{\mathcal{S}\mathcal{R}}}(D-\delta-d_{\mathcal{S}\mathcal{R}})^\alpha}{c_{\mathcal{R}\mathcal{D}_q}P_{\mathcal{R}_q}}\Big)^\mathcal{B}\Big]^{\frac{-1}{\mathcal{B}}}$. With the pseudoconcavity of objective function and convexity of $C1$, $C2,\widehat{C3}$, the Karush-Kuhn-Tucker (KKT) point of (P1) yields its global optimal solution.
Further, the Lagrangian function of (P1) by associating the Lagrange multiplier $\lambda$ with $\widehat{C3}$ and considering $C1$ and $C2$ implicit, can be defined by: 
\begin{equation}\label{LagFun1} 
\textstyle\mathcal{L}_1=\mathrm \prod_{q=1}^{n}\left(1+\mathbb{E}[{\min\{\gamma_{\mathcal{S}\mathcal{R}_q},\gamma_{\mathcal{R}\mathcal{D}_q}\}}]\right)-\lambda\mathcal{J}, 
\end{equation}
where $\mathcal{J}\triangleq\big(\sum_{q=1}^{n}(P_{\mathcal{S}_q}+P_{\mathcal{R}_q})-P_B\big)$.
On simplifying the KKT conditions $\big[\frac{\partial \mathcal{L}_1}{\partial P_{\mathcal{S}_q}}=0$, $\frac{\partial \mathcal{L}_1}{\partial P_{\mathcal{R}_q}}=0$, $\lambda\mathcal{J}=0$,\,$C1$, $C2$, $\widehat{C3}$, and $\lambda \geq 0\big]$,  we get a system of $(2n+2)$ equations represented by~\eqref{OptimizationA},~\eqref{OptimizationB},~\eqref{OptimizationC} and $\mathcal{J}$, to be solved  $\{P_{\mathcal{S}_q},P_{\mathcal{R}_q}\}_{q=1}^n,d_{\mathcal{S}\mathcal{R}}$ and $\lambda$. Variables $Q_q,T_q , V_q,\forall q\le n,$ in  \eqref{eq:KKTJ} are defined below.
\begin{subequations}\label{three_var}
	\begin{eqnarray}\label{Qq_var}
	\hspace{-3mm}&Q_q\!=\!\frac{\beta}{\lambda N_q \Delta f}\bigg[\!\!\left(\!\!\frac{c_{\mathcal{S}\mathcal{R}_q}}{a_q^{d_{\mathcal{S}\mathcal{R}}}d_{\mathcal{S}\mathcal{R}}^{\alpha}}\!\right)^{\!\!\frac{\mathcal{B}}{\mathcal{B}+1}}\hspace{-1.5mm}+\hspace{-1mm}\left(\!\!\frac{c_{\mathcal{R}\mathcal{D}_q} a_q^{\hspace{-1mm}-(D-\delta-d_{\mathcal{S}\mathcal{R}})}}{ (D-\delta-d_{\mathcal{S}\mathcal{R}})^{\alpha}}\!\right)^{\!\!\frac{\mathcal{B}}{\mathcal{B}+1}}\bigg]^{\!\!\frac{\mathcal{B}+1}{\mathcal{B}}}\!\!\!\!,
	\end{eqnarray}
	\begin{align}\label{Tq_var}\textstyle T_q\hspace{-0.5mm}=\hspace{-0.5mm} (\beta c_{\mathcal{S}\mathcal{R}_q}\hspace{-0.5mm}[\lambda N_q \Delta f a_q^{d_{\mathcal{S}\mathcal{R}}}d_{\mathcal{S}\mathcal{R}}^\alpha]^{\hspace{-0.5mm}-1}\hspace{-0.5mm})^{\frac{\mathcal{B}}{\mathcal{B}+1}}\hspace{-1mm}-\hspace{-1mm}1,	
	\end{align}
	\begin{align}\label{Vq_var}
	V_q=&\, (c_{\mathcal{S}\mathcal{R}_q}P_{{\mathcal{S}_q}} a_q^{D-\delta-d_{\mathcal{S}\mathcal{R}}} (D-\delta-d_{\mathcal{S}\mathcal{R}})^\alpha)^\mathcal{B}\nonumber\\
	&\,+(c_{\mathcal{R}\mathcal{D}_q}P_{\mathcal{R}_q}a_q^{d_{\mathcal{S}\mathcal{R}}}d_{\mathcal{S}\mathcal{R}}^\alpha)^\mathcal{B}.	
	\end{align}
\end{subequations}
\color{black} As it is cumbersome to solve system of $(2n+2)$  equations for large value of $n$ to ensure the equivalence of problems $\text{(P0)}$ and $\text{(P1)}$, we next propose a novel low-complexity approximation.
\begin{figure*}[!t]
	\begin{subequations}\label{eq:KKTJ}
		\begin{equation}\label{OptimizationA}
		\begin{aligned}
		P_{\mathcal{S}_q}=P_{\mathcal{R}_q}c_{\mathcal{R}\mathcal{D}_q}\left[c_{\mathcal{S}\mathcal{R}_q}a_q^{D-\delta-2d_{\mathcal{S}\mathcal{R}}}\left(\left(D-\delta\right)d_{\mathcal{S}\mathcal{R}}^{-1}-1\right)^\alpha\right]^{-1}\left[\left(Q_q\beta c_{\mathcal{S}\mathcal{R}_q}\left[N_q \lambda \Delta f a_q^{d_{\mathcal{S}\mathcal{R}}}d_{\mathcal{S}\mathcal{R}}^\alpha\right]^{-1}\right)^\frac{\mathcal{B}}{\mathcal{B}+1}-1\right]^\frac{1}{\mathcal{B}}
		\end{aligned} 
		\end{equation}
		 
		\begin{eqnarray}\label{OptimizationB}
		P_{\mathcal{R}_q}=P_{\mathcal{S}_q}c_{\mathcal{S}\mathcal{R}_q}a_q^{D-\delta-2d_{\mathcal{S}\mathcal{R}}}\left(\left(D-\delta\right)d_{\mathcal{S}\mathcal{R}}^{-1}-1\right)^\alpha c_{\mathcal{R}\mathcal{D}_q}^{-1}\left[\left(Q_q\beta c_{\mathcal{R}\mathcal{D}}\left[N_q \lambda \Delta f a_q^{D-d_{\mathcal{S}\mathcal{R}}-\delta}\left(D-d_{\mathcal{S}\mathcal{R}}-\delta\right)^\alpha\right]^{-1}\right)^\frac{\mathcal{B}}{\mathcal{B}+1}-1\right]^\frac{1}{\mathcal{B}}
		\end{eqnarray}   
		 
		\begin{eqnarray}\label{OptimizationC} 
		\sum_{q=1}^{n}\beta c_{\mathcal{S}\mathcal{R}_q}c_{\mathcal{R}\mathcal{D}_q}P_{\mathcal{S}_q}P_{\mathcal{R}_q}\left[N_q \Delta f\right]^{-1} V_q^{\frac{-\mathcal{B}}{\mathcal{B}+1}}\hspace{-1mm}\left(c_{\mathcal{R}\mathcal{D}_q}P_{\mathcal{R}_q}a_q^{d_{\mathcal{S}\mathcal{R}}}d_{\mathcal{S}\mathcal{R}}^\alpha\right)^\mathcal{B}\hspace{-1mm}\left[T_q\left(\ln a_q\hspace{-0.8mm} +\hspace{-0.8mm}\alpha \left(D\hspace{-0.5mm}-\hspace{-0.5mm}\delta\hspace{-0.5mm}-\hspace{-0.5mm}d_{\mathcal{S}\mathcal{R}}\right)^{-1}\right)\hspace{-0.8mm}-\hspace{-0.8mm}\left(\ln a_q\hspace{-0.5mm} +\hspace{-0.5mm}\alpha d_{\mathcal{S}\mathcal{R}}^{-1}\hspace{-0.5mm}\right)\right]\hspace{-0.5mm}=\hspace{-0.5mm}0\!\!\! 
		\end{eqnarray}
	\end{subequations}	
	
	\hrulefill
\end{figure*}

\section{Low Complexity Approximation Algorithm}\label{section4}
This proposed algorithm  decoupling the joint optimization into individual PA and RP problems, can be summarized into three main steps as discussed  in following three subsections.
\subsection{Optimal PA (OPA) within a sub-band for a given RP}\label{sec:PA-a}
 For a given RP, we first distribute the power budget $P_{t_q}$ for sub-band $q$ between  $P_{{\mathcal{S}_q}}$ and $P_{{\mathcal{R}_q}}$ to maximize $\overline{\gamma}_q$. As with $P_{\mathcal{R}_q}=P_{t_q}-P_{\mathcal{S}_q}$,   $\overline{\gamma}_q$ is concave in $P_{\mathcal{S}_q}$, optimal values $P_{\mathcal{S}_q}^*$ and $P_{\mathcal{R}_q}^*=\mathcal{Z}_qP_{\mathcal{S}_q}^*$ are obtained on solving $\textstyle\frac{\partial\overline{\gamma}_q}{\partial P_{\mathcal{S}_q}}=0$, where
\begin{equation}\label{Zq_var} 
	\!\textstyle\mathcal{Z}_q \triangleq  (\hspace{-0.5mm}c_{\mathcal{S}\mathcal{R}_q}\hspace{-0.1mm}a_q^{D\hspace{-0.1mm}-\hspace{-0.1mm}\delta\hspace{-0.1mm}-\hspace{-0.1mm}d_{\mathcal{S}\mathcal{R}}}(D\hspace{-0.5mm}-\hspace{-0.5mm}\delta\hspace{-0.5mm}-\hspace{-0.5mm}d_{\mathcal{S}\mathcal{R}}\hspace{-0.5mm})^\alpha[c_{\mathcal{R}\mathcal{D}_q}\hspace{-0.5mm}a_q^{\hspace{-0mm}d_{\mathcal{S}\mathcal{R}}}{d_{\mathcal{S}\mathcal{R}}^\alpha}]^{-1}\hspace{-0.5mm})^\frac{\mathcal{B}}{\mathcal{B}+1}\hspace{-1mm} 
\end{equation}

\noindent\color{black} Here, note that $P_{\mathcal{S}_q}\gtrless P_{\mathcal{R}_q}$ as determined by $\mathcal{Z}_q\lessgtr1$ depends on the relative received SNRs over $\mathcal{S}\mathcal{R}$ and $\mathcal{R}\mathcal{D}$ links. 

\begin{figure*}[!t]
	\centering
	\subfigure[]{\includegraphics[width=2.3in]{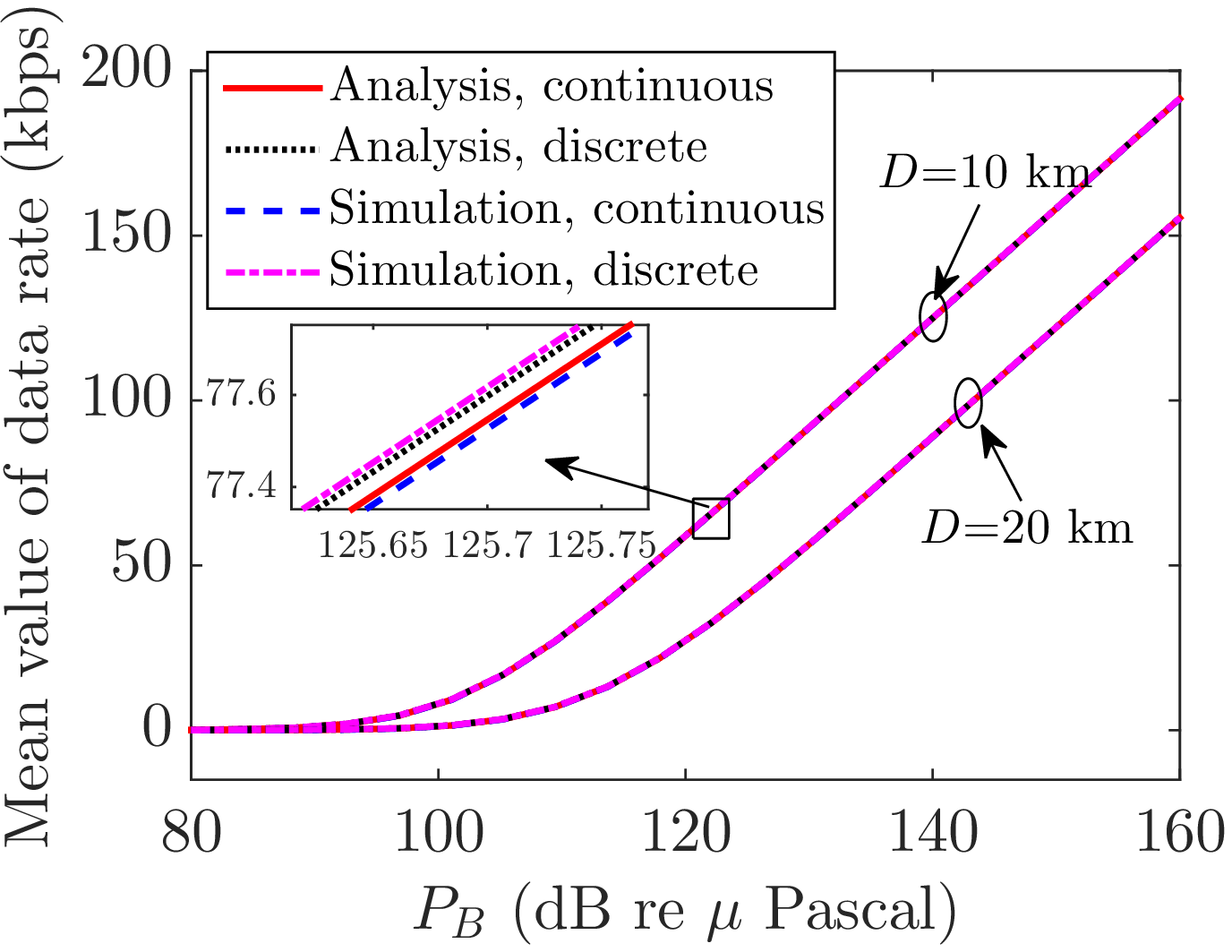}} 
	\subfigure[]{\includegraphics[width=2.3in]{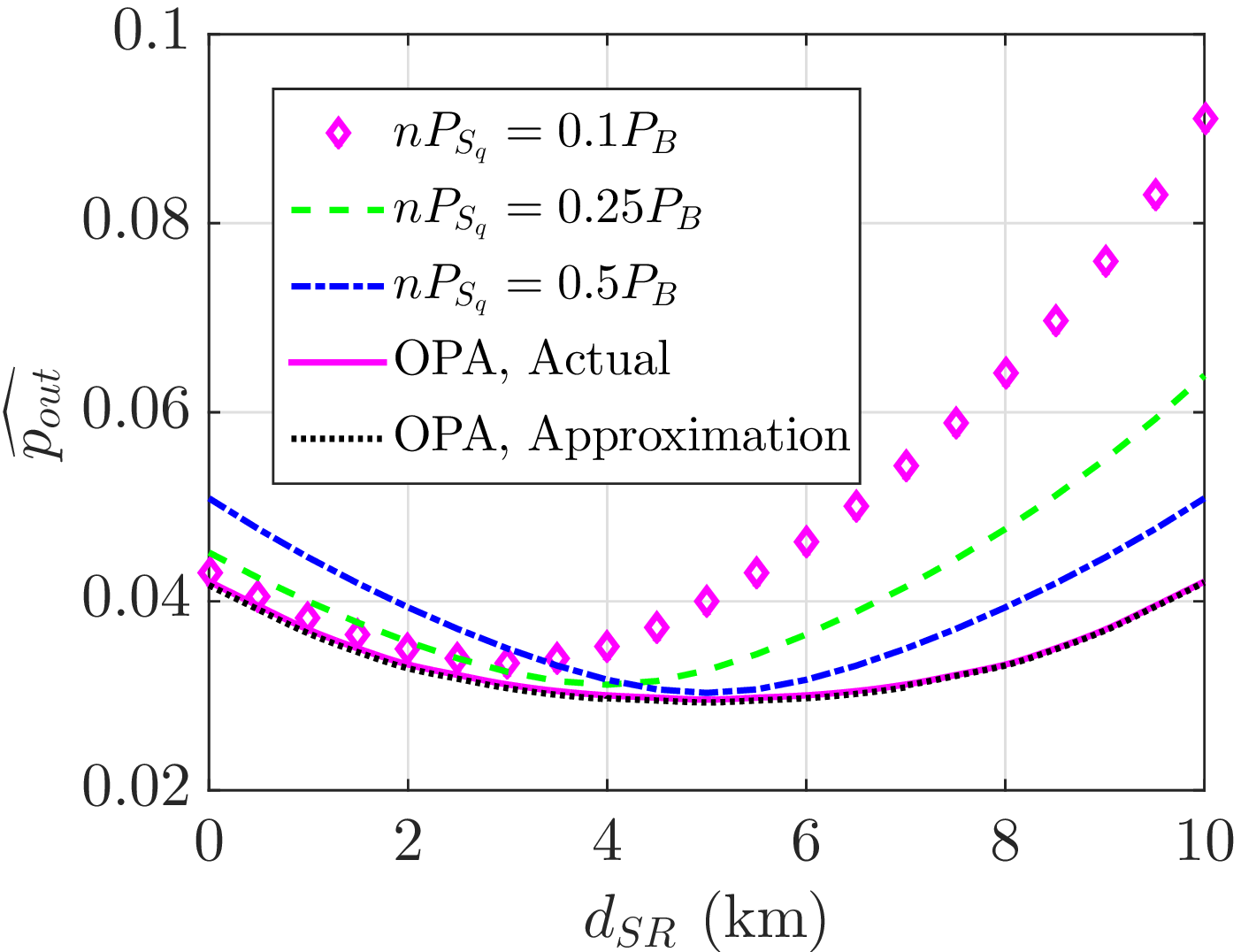}} 
	\subfigure[]{\includegraphics[width=2.3in]{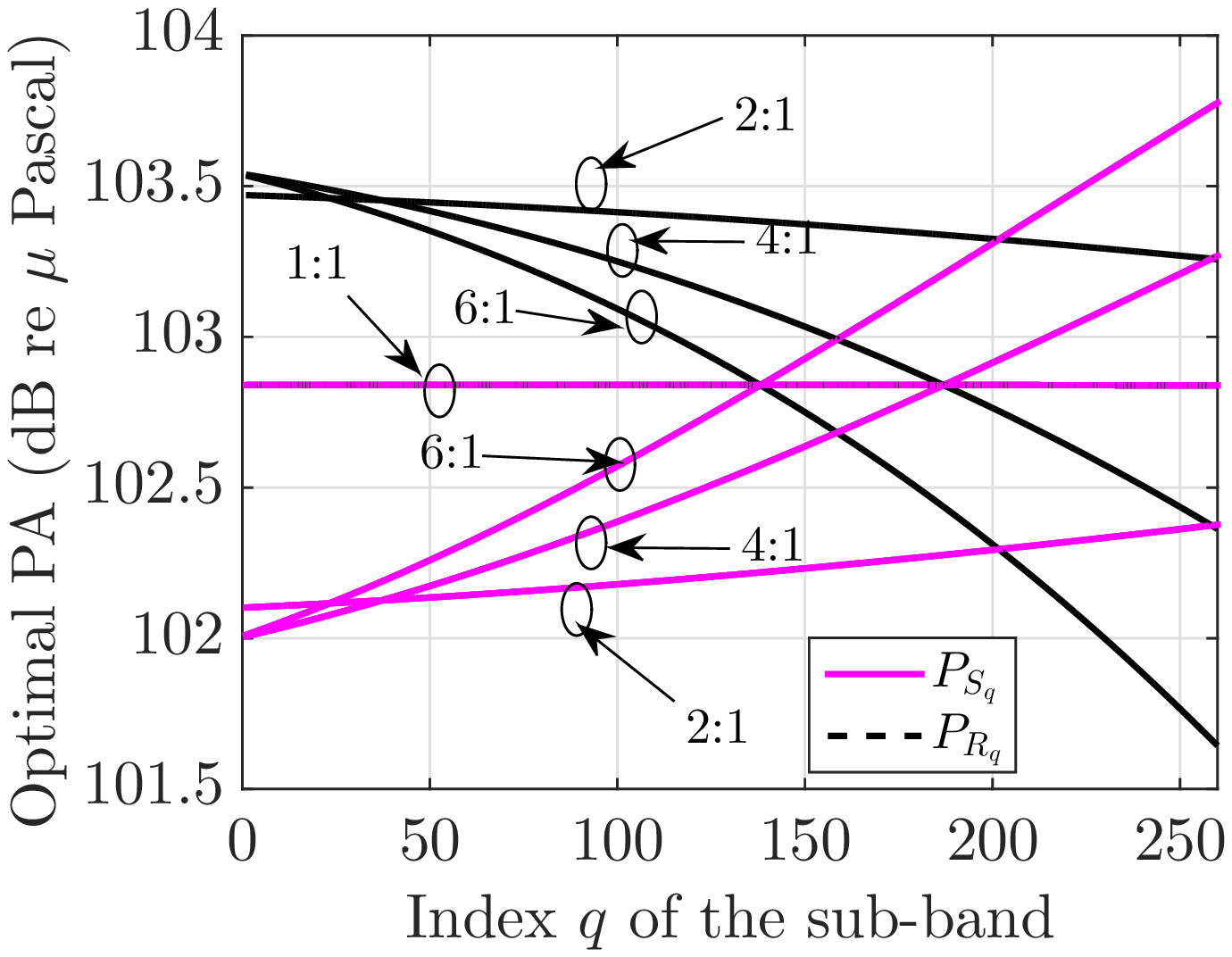}}
	\caption{\small Validation of analysis and insights on OPA and ORP with varying system parameters. (a) Variation of expected data rate with $P_B$ in continuous and discrete domains. (b) Variation of $\widehat{p_{out}}$ with $d_{\mathcal{S}\mathcal{R}}$ with FPA. (c) Variation of OPA across sub-bands with $c_{\mathcal{S}\mathcal{R}}(f):c_{\mathcal{R}\mathcal{D}}(f)$.}
	\label{fig:validation}
\end{figure*} 

\subsection{OPA to each sub-band for a given $\{P_{\mathcal{R}_q}\}_{q=1}^n$ and $d_{\mathcal{S}\mathcal{R}}$}
Using this derived relationship $P_{\mathcal{R}_q}=\mathcal{Z}_qP_{\mathcal{S}_q}$, we can eliminate $\{P_{\mathcal{R}_q}\}_{q=1}^n$ in~\eqref{LagFun1} and hence obtain an updated Lagrangian $\mathcal{L}_2$ which is a function of only $n+2$ variables:
\begin{equation}\label{LangFun2}
 \hspace{-0.5mm}\textstyle\mathcal{L}_2\hspace{-0.5mm}=\hspace{-0.5mm}\prod_{q=1}^{n}\hspace{-0.5mm}\left(1\hspace{-0.5mm}+\hspace{-0.5mm}P_{\mathcal{S}_q}[\mathcal{K}_q]^{-1}\right)\hspace{-0.5mm}-\hspace{-0.5mm}\lambda\hspace{-0.25mm}\big(\hspace{-0.5mm}\textstyle\sum_{q=1}^{n}P_{\mathcal{S}_q}(1\hspace{-0.5mm}+\hspace{-0.5mm}\mathcal{Z}_q)\hspace{-0.5mm}-\hspace{-0.5mm}P_B\big), 
\end{equation}
where $\mathcal{K}_q=N_q \Delta f a_q^{d_{\mathcal{S}\mathcal{R}}}{d_{\mathcal{S}\mathcal{R}}^\alpha}\Big(1+\mathcal{Z}_q^\frac{1}{\mathcal{B}}\Big)^\frac{1}{\mathcal{B}}[\beta c_{\mathcal{S}\mathcal{R}_q}]^{-1}$. Now to obtain the optimal $\{P_{\mathcal{S}_q}\}_{q=1}^n$ and $\lambda$ for given  $d_{\mathcal{S}\mathcal{R}}$ and $P_{\mathcal{R}_q}=\mathcal{Z}_qP_{\mathcal{S}_q}\;\forall q$, the corresponding KKT conditions are:
\begin{subequations}
	\begin{eqnarray}\label{condA}
	\begin{aligned}
	\textstyle\frac{\partial \mathcal{L}_2}{\partial P_{\mathcal{S}_q}}=\frac{1}{\mathcal{K}_q}\textstyle\prod_{j=1,j\neq q}^{n}\left(1+\frac{P_{\mathcal{S}_j}}{\mathcal{K}_j}\right)-\lambda (1+\mathcal{Z}_q)=0,
	\end{aligned} 
	\end{eqnarray}
	\begin{eqnarray}\label{condB}
	\begin{aligned}
	\textstyle \lambda\left(\sum_{q=1}^{n}(1+\mathcal{Z}_q)P_{\mathcal{S}_q}-P_B\right)=0.
	\end{aligned}
	\end{eqnarray}
\end{subequations}

As for $\lambda^*=0$,~\eqref{condA} cannot be satisfied, we note that $\lambda^*>0$.  On solving~\eqref{condA} and~\eqref{condB},  $\{P_{\mathcal{S}_q}^*\}_{q=1}^n$ and $\lambda^*$ are obtained as: 
\begin{subequations}
	\begin{eqnarray}\label{eq:OPA-S}
	\begin{aligned}
	\textstyle P_{\mathcal{S}_q}^*\triangleq \frac{P_B+\textstyle\sum_{j=1}^{n}(1+\mathcal{Z}_j)\mathcal{K}_j-(1+\mathcal{Z}_q)\mathcal{K}_q}{n(1+\mathcal{Z}_q)},\quad
	\end{aligned} 
	\end{eqnarray}
	\begin{eqnarray}\label{eq:opt-lam}
	\begin{aligned}
	\textstyle \lambda^*\triangleq(1+\mathcal{Z}_1)(\mathcal{K}_1+P_{\mathcal{S}_1})^{n-1}\textstyle\prod_{j=1}^{n}\frac{1}{\mathcal{K}_j(1+\mathcal{Z}_j)}.
	\end{aligned}
	\end{eqnarray}
\end{subequations}
\color{black} Further, as for practical system parameter values in UANs, $P_B\gg \textstyle\sum_{j=1}^{n}(1+\mathcal{Z}_j)\mathcal{K}_j-n(1+\mathcal{Z}_q)\mathcal{K}_q$, we note that $ P_{\mathcal{S}_q}^* \approx P_B[n(1+\mathcal{Z}_q)]^{-1}$.
\textit{Hence, this approximation along with~\eqref{eq:OPA-S} and $P_{\mathcal{R}_q}=\mathcal{Z}_qP_{\mathcal{S}_q}$ provide novel insights on OPA across different sub-bands as a function of $f_q$ and RP $d_{\mathcal{S}\mathcal{R}}$.}

\subsection{Optimal Positioning of Relay for the Obtained OPA}

Using~\eqref{eq:OPA-S} and~\eqref{eq:opt-lam} in~\eqref{LangFun2}, $\mathcal{L}_2$ having $n+2$  variables gets reduced to a single variable Lagrangian $\mathcal{L}_3$ after writing $\{P_{\mathcal{S}_q}\}_{q=1}^n$ and $\lambda$ as functions of RP $d_{\mathcal{S}\mathcal{R}}$. Thus, we get optimal RP  $d_{\mathcal{S}\mathcal{R}}^*$ by solving $\frac{\partial \mathcal{L}_3}{\partial d_{\mathcal{S}\mathcal{R}}}=0$, and then the OPA $P_{\mathcal{S}_q}^*$ by substituting $d_{\mathcal{S}\mathcal{R}}^*$ in~\eqref{eq:OPA-S} and $P_{\mathcal{R}_q}^*$ by $P_{\mathcal{R}_q}^*=\mathcal{Z}_qP_{\mathcal{S}_q}^*$. Here, it is worth noting that, \textit{regardless of value of $n\gg1$}, \textit{we just need to solve one single variable equation $\frac{\partial \mathcal{L}_3}{\partial d_{\mathcal{S}\mathcal{R}}}=0$ to obtain the tight approximation to the joint global-optimal solution as obtained by solving the system of $(2n+2)$ equations.} This in turn yields huge reduction in computational time complexity.

%\begin{equation}\label{eq6}
%\begin{aligned}
%&\text{P1:}&&\\
%&\hspace{-4mm} \underset{S_\mathcal{S}(f_i),S_\mathcal{R}(f_i),d_{\mathcal{S}\mathcal{R}_i}}{\text{minimize}} \underset{i}{\text{max}}
%&&\hspace{-2mm}\textbf{Pr}\left(\mathrm \frac{1}{2}\log_2(1+\min\{\gamma_{\mathcal{S}\mathcal{R}_i}(f_i),\gamma_{\mathcal{R}\mathcal{D}_i}(f_i)\})\leq \frac{\mathcal{R}}{n}\right) \\
%&\hspace{-2mm} \text{subject to}
%&&\hspace{-6mm}C1, C2,
%\end{aligned}
%\end{equation}

\section{Numerical Results}\label{sec_results}
The default experimental parameters are as follows. Operating frequency range is between $5$ to $15$ kHz~\cite{stoj02}, $c_{\mathcal{S}\mathcal{R}}(f)=c_{\mathcal{R}\mathcal{D}}(f)$ which is assumed to be constant over entire operating bandwidth~\cite{babu10}, $D=10$ km, $d_{\mathcal{S}\mathcal{R}}=5$ km, $n=260$, $r=1$ kbps, $K=3.01$ dB, $\alpha=1.5$, and $P_B=100$ dB re $\mu$ Pascal. 

First we validate the analysis by plotting the mean value of data rate in both continuous and discrete frequency domains (with $n=260$) in Fig.~\ref{fig:validation}(a). A percentage error of $\le0.02\%$ between the analytical and simulation results in each case validates that with $n\ge260$, $\widehat{p_{out}}$ closely matches $p_{out}$. Further via Fig.~\ref{fig:validation}(b), minimum  $\widehat{p_{out}}$  obtained using the low complexity approximation algorithm (cf. Section~\ref{section4}) differs  by less than $0.032\%$ from the global minimum value as returned by solving $(2n+2)$ equations for obtaining solution of (P1).

Next we get insights on OPA and optimal RP (ORP). In Fig.~\ref{fig:validation}(b), the performance of different fixed PA (FPA) schemes is compared against OPA for varying RPs. If total PA $nP_{\mathcal{S}_q}$ at $\mathcal{S}$ in FPA increases, the minimum $\widehat{p_{out}}$ is obtained when $\mathcal{R}$ is located near $\mathcal{D}$. The uniform PA (UPA), having $P_{\mathcal{S}_q}=P_{\mathcal{R}_q}={P_B}/(2n)$ $\forall$ $q$, achieves nearly the same global minimum value of $\widehat{p_{out}}$ approximately at same point $d_{\mathcal{S}\mathcal{R}}=0.5D$. Because on using $c_{\mathcal{S}\mathcal{R}}(f)=c_{\mathcal{R}\mathcal{D}}(f)$ and $d_{\mathcal{S}\mathcal{R}}=0.5D$ in~\eqref{eq:OPA-S}, $\mathcal{Z}_q=1$ $\forall$ $q$, and as a result OPA is independent of center frequencies. \textit{Thus, for symmetric channels, i.e., $c_{\mathcal{S}\mathcal{R}}(f)=c_{\mathcal{R}\mathcal{D}}(f)$, OPA on sub-bands is uniform regardless of the values of $\{f_q\}_{q=1}^n$, as also evident from Fig.~\ref{fig:validation}(c)}. However, in practice for asymmetric $\mathcal{S}\mathcal{R}$ and $\mathcal{R}\mathcal{D}$ links, we need to obtain OPA using proposed algorithm.

The variation of OPA along the sub-bands vary with different channel gains for $\mathcal{S}\mathcal{R}$ and $\mathcal{R}\mathcal{D}$ link is shown in Fig.~\ref{fig:validation}(c). When $c_{\mathcal{S}\mathcal{R}}(f):c_{\mathcal{R}\mathcal{D}}(f)=2:1$, $\mathcal{S}$ requires lower PA and optimal RP is nearer to $\mathcal{D}$, because channel gain of $\mathcal{S}\mathcal{R}$ link is higher. But for $c_{\mathcal{S}\mathcal{R}}(f):c_{\mathcal{R}\mathcal{D}}(f)=4:1$ and $6:1$, initially the OPA is lower at $\mathcal{S}$ followed by an inversion taking place due to $Z_q<1$ at $q\geq 138$ and $\geq 188$, respectively, because the relative attenuation $\frac{a_q^{d_{\mathcal{S}\mathcal{R}}}d_{\mathcal{S}\mathcal{R}}^\alpha}{a_q^{D-\delta-d_{\mathcal{S}\mathcal{R}}}{(D-\delta-d_{\mathcal{S}\mathcal{R}})}^\alpha}$ dominates over the relative expected gain of  $\frac{c_{\mathcal{S}\mathcal{R}}(f)}{c_{\mathcal{R}\mathcal{D}}(f)}$ of $\mathcal{S}\mathcal{R}$ to $\mathcal{R}\mathcal{D}$ link (cf. Section~\ref{sec:PA-a}). \textit{Therefore, OPA along a sub-band over $\mathcal{S}\mathcal{R}$ and $\mathcal{R}\mathcal{D}$ link depends on dominance of relative gain of fading channels over relative channel attenuation, and vice versa.} 

\begin{figure} 
	\centering  \includegraphics[width=3.5in]{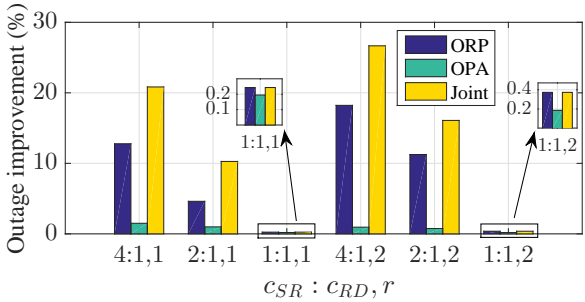}
	\caption{\small Percentage improvement achieved by different proposed optimization schemes over FPA for different $[c_{\mathcal{S}\mathcal{R}}(f):c_{\mathcal{R}\mathcal{D}}(f), r]$.}    \label{fig:compare}
\end{figure}

Finally, we compare the outage performance of the three optimization schemes, (i) ORP with UPA, (ii) OPA with $d_{\mathcal{S}\mathcal{R}}=0.5D$, and (iii) joint PA and RP, against a fixed benchmark scheme with UPA and $d_{\mathcal{S}\mathcal{R}}=0.5D$ (cf. Fig.~\ref{fig:compare}). The average percentage improvement provided by ORP, OPA, and joint optimization schemes are  $15.5\%$, $1.2\%$, and $23.85\%$ respectively for $c_{\mathcal{S}\mathcal{R}}(f):c_{\mathcal{R}\mathcal{D}}(f)=4:1$, and $0.31\%$, $0.19\%$, and $0.31\%$ for $c_{\mathcal{S}\mathcal{R}}(f):c_{\mathcal{R}\mathcal{D}}(f)=1:1$. Also, the same is true for reverse ratio, i.e., $c_{\mathcal{S}\mathcal{R}}(f):c_{\mathcal{R}\mathcal{D}}(f)=1:2 $ and $1:4$. \textit{Thus, higher the asymmetry in channel gains of $\mathcal{S}\mathcal{R}$ and $\mathcal{R}\mathcal{D}$ links, higher is the percentage improvement in performance and the ORP is a better semi-adaptive scheme than OPA.}

\section{Concluding Remarks} 
We jointly optimized PA and RP to minimize outage probability. After proving the global optimality of the problem, we also propose an efficient, tight approximation algorithm which substantially reduces the complexity in calculation. In general, the numerically validated proposed analysis and joint optimization have been shown to provide more than $10\%$ outage improvement over the fixed benchmark scheme. Though this performance enhancement depends on the  $\mathcal{S}\mathcal{R}$ and $\mathcal{R}\mathcal{D}$ channel gains, the cost incurred in practically realizing them is negligible due to the proposed low complexity design.

%\section{Concluding Remarks}     
%\label{sec_conclusion}
%Write the conclusion here.
 
\appendices
\setcounter{equation}{0}
\setcounter{figure}{0}
\renewcommand{\theequation}{A.\arabic{equation}}
\renewcommand{\thefigure}{A.\arabic{figure}} 

\section{}
\subsection{Proof of Pseudoconvexity of $p_{out}$ in $S_\mathcal{S}$, $S_\mathcal{R}$, $d_{\mathcal{S}\mathcal{R}}$}\label{AppA} 	
From~\eqref{out_exp}, we notice that the outage probability $p_{out}$ can be observed as the CDF of the random rate $\mathfrak{R}\triangleq\int_{0}^{B_W}\frac{1}{2}\log_2(1+\min\{\gamma_{\mathcal{S}\mathcal{R}}(f),\gamma_{\mathcal{R}\mathcal{D}}(f)\})\text{d}f$.  It is clear that $\mathfrak{R}$ depends on the end-to-end SNR $\gamma=\min\{\gamma_{\mathcal{S}\mathcal{R}}(f),\gamma_{\mathcal{R}\mathcal{D}}(f)\}$, whose expectation as obtained using the relationship $\textbf{Pr}[\gamma>x]=\textbf{Pr}[\gamma_{\mathcal{S}\mathcal{R}}>x]\textbf{Pr}[\gamma_{\mathcal{R}\mathcal{D}}>x]$ in~\eqref{ccdf}, is given by $\overline{\gamma}= \frac{\overline{\gamma}_{\mathcal{S}\mathcal{R}}\hspace{1mm}\overline{\gamma}_{\mathcal{R}\mathcal{D}}}{[\overline{\gamma}_{\mathcal{S}\mathcal{R}}^\mathcal{B}+\overline{\gamma}_{\mathcal{R}\mathcal{D}}^\mathcal{B}]^\frac{1}{\mathcal{B}}}$. After using the definitions for $\overline{\gamma}_{ij}$ (as given in Section~\ref{section2}), we obtain:
\begin{equation}\label{meanSNR} 
\overline{\gamma}\!=\!\textstyle\frac{\beta}{N(f)}\Big[\Big(\frac{a(f)^{d_{\mathcal{S}\mathcal{R}}}d_{\mathcal{S}\mathcal{R}}^\alpha}{c_{\mathcal{S}\mathcal{R}}S_{S}(f)}\Big)^\mathcal{B}+\Big(\frac{a(f)^{D-\delta-d_{\mathcal{S}\mathcal{R}}}(D-\delta-d_{\mathcal{S}\mathcal{R}})^\alpha}{c_{\mathcal{R}\mathcal{D}}S_{R}(f)}\Big)^\mathcal{B}\Big]^{\frac{-1}{\mathcal{B}}} 
\end{equation}
As the distribution of $\mathfrak{R}$ depends on SNR $\gamma$, using the joint pseudoconcavity of $\overline{\gamma}$ as proved in Appendix~\ref{AppB}, it can be shown that the expectation $\overline{\mathfrak{R}}$ of $\mathfrak{R}$ is also jointly pseudoconcave in  $S_\mathcal{S}(f)$, $S_\mathcal{R}(f)$, and $d_{\mathcal{S}\mathcal{R}}$. The latter holds because the affine and logarithmic transformation along with integration preserve the pseudoconcavity of the positive pseudoconcave function $\overline{\gamma}$~\cite{Baz06},~\cite[App. C]{mish17}. Finally, using the property that the CDF is a monotonically decreasing function of the expectation  of the underlying random variable~\cite[Theorem 1]{mish18}, we observe that $p_{out}$, which holds a similar CDF and expectation relationship with $\overline{\mathfrak{R}}$, is jointly pseudoconvex~\cite{Baz06}.
\subsection{Proof of Pseudoconcavity of $\overline{\gamma}$ in $S_\mathcal{S}$, $S_\mathcal{R}$, and $d_{\mathcal{S}\mathcal{R}}$}\label{AppB}
 The bordered Hessian matrix $B_H(\overline{\gamma})$ for $\overline{\gamma}$ is given by:	
	\begin{eqnarray}\label{eq:ApB1}
	B_H(\overline{\gamma})=\left[\hspace{-0.5mm} \begin{array}{cccc}
	0 & \frac{\partial \overline{\gamma}}{\partial {S_\mathcal{S}}} & \frac{\partial \overline{\gamma}}{\partial {S_\mathcal{R}}} & \frac{\partial \overline{\gamma}}{\partial d_{\mathcal{S}\mathcal{R}}} \\ \frac{\partial \overline{\gamma}}{\partial {S_\mathcal{S}}} &  \frac{\partial^2 \overline{\gamma}}{\partial {S_\mathcal{S}}^2} & \frac{\partial^2 \overline{\gamma}}{\partial {S_\mathcal{S}} \partial {S_\mathcal{R}}} & \frac{\partial^2 \overline{\gamma}}{\partial {S_\mathcal{S}} \partial {d_{\mathcal{S}\mathcal{R}}}}\\ \frac{\partial \overline{\gamma}}{\partial {S_\mathcal{R}}} & \frac{\partial^2 \overline{\gamma}}{\partial {S_\mathcal{R}} \partial {S_\mathcal{S}}} & \frac{\partial^2 \overline{\gamma}}{\partial {S_\mathcal{R}}^2} & \frac{\partial^2 \overline{\gamma}}{\partial {S_\mathcal{R}} \partial d_{\mathcal{S}\mathcal{R}}} \\ \frac{\partial \overline{\gamma}}{\partial d_{\mathcal{S}\mathcal{R}}} & \frac{\partial^2 \overline{\gamma}}{\partial d_{\mathcal{S}\mathcal{R}}  \partial {S_\mathcal{S}} } & \frac{\partial^2 \overline{\gamma}}{\partial d_{\mathcal{S}\mathcal{R}}  \partial {S_\mathcal{R}} } & \frac{\partial^2 \overline{\gamma}}{\partial d_{\mathcal{S}\mathcal{R}}^2} \end{array}    \right]
	\end{eqnarray} 
	From \eqref{eq:ApB1}, the joint pseudoconcavity of $\overline{\gamma}$  in $S_\mathcal{S}(f)$, $S_\mathcal{R}(f)$, and $d_{\mathcal{S}\mathcal{R}}$ is proved next by showing that the determinant of $3\times3$ leading principal submatrix of $B_H(\overline{\gamma})$, denoted by $\mathfrak{L}$, is positive, and the determinant of $B_H(\overline{\gamma})$ is negative~\cite{Baz06}.
	\begin{subequations}
	\begin{align}\label{L_var}
		\begin{aligned}
		\hspace{-3mm}|\mathfrak{L}|=(1+\mathcal{B})Y_1^\mathcal{B}Y_2^\mathcal{B}(Y_1^\mathcal{B}+Y_2^\mathcal{B})^{-3-\frac{3}{\mathcal{B}}}(S_\mathcal{S}S_\mathcal{R})^{-2}>0,\!\!
	\end{aligned}
	\end{align} 
	\begin{align}\label{Bh_var}
		 \hspace{-1mm}|B_H(\overline{\gamma})&|\hspace{-1mm}=\hspace{-1mm}-\{Y_1^\mathcal{B}Y_2^\mathcal{B}(\hspace{-0.5mm}Y_1^\mathcal{B}\hspace{-1mm}+\hspace{-1mm}Y_2^\mathcal{B}\hspace{-0.5mm})^{\hspace{-0.5mm}-2\hspace{-0mm}-\hspace{-0mm}\frac{3}{\mathcal{B}}}(d_{\mathcal{S}\mathcal{R}}(\hspace{-0.5mm} D\hspace{-1mm}-\hspace{-1mm}\delta\hspace{-1mm}-\hspace{-1mm}d_{\mathcal{S}\mathcal{R}}\hspace{-0.5mm})\nonumber\\
		 &\;\times\hspace{-0.5mm} S_\mathcal{S}S_\mathcal{R})^{\hspace{-0.5mm}-2}\hspace{-0.5mm}\}\{\hspace{-0.5mm}\alpha(\alpha\hspace{-1mm}-\hspace{-1mm}1)(1\hspace{-1mm}+\hspace{-1mm}\mathcal{B}\hspace{-0mm})((\hspace{-0.5mm}D\hspace{-1mm}-\hspace{-1mm}\delta\hspace{-0.5mm})Y_1\hspace{-1mm}-\hspace{-1mm}d_{\mathcal{S}\mathcal{R}}(Y_1\hspace{-1mm}+\hspace{-1mm}Y_2))^2\nonumber\\
		 &\;+\hspace{-0.5mm}\alpha(\mathcal{B}(\alpha\hspace{-1mm}-\hspace{-1mm}1)\hspace{-1mm}-\hspace{-1mm}1)(D\hspace{-1mm}-\hspace{-1mm}\delta)^2Y_1Y_2\}\hspace{-1mm}+\hspace{-1mm}2\alpha d_{\mathcal{S}\mathcal{R}}(D\hspace{-1mm}-\hspace{-1mm}\delta\hspace{-1mm}-\hspace{-1mm}d_{\mathcal{S}\mathcal{R}})\nonumber\\
		 &\;\times \ln a\{(1+\mathcal{B})d_{\mathcal{S}\mathcal{R}}Y_2^2+(1+\mathcal{B})(D-\delta-d_{\mathcal{S}\mathcal{R}})Y_1^2\nonumber\\
		 &\;+\hspace{-0.5mm}(\mathcal{B}\hspace{-1mm}-\hspace{-1mm}1)(D\hspace{-1mm}-\hspace{-1mm}\delta)Y_1Y_2\} \hspace{-0.5mm}+\hspace{-0.5mm} d^2(D\hspace{-1mm}-\hspace{-1mm}\delta\hspace{-1mm}-\hspace{-1mm}d_{\mathcal{S}\mathcal{R}})^2(\ln a)^2\hspace{-0.5mm}\{(Y_1\hspace{-1mm}\nonumber\\
		 &\;-\hspace{-1mm}Y_2)^2+\hspace{-1mm}\mathcal{B}(Y_1\hspace{-1mm}+\hspace{-1mm}Y_2)^2\}\hspace{-1mm}<\hspace{-1mm}0, \hspace{1mm}\forall\{(\alpha\hspace{-1mm}>\hspace{-1mm}1)\hspace{-0.5mm}\wedge\hspace{-0.5mm} (\mathcal{B}\hspace{-1mm} > \hspace{-1mm}1)\}\!\!\!\!\!\!\!\!
	\end{align} 
    \end{subequations}
    Here $Y_1\triangleq\frac{a^{d_{\mathcal{S}\mathcal{R}}}d_{\mathcal{S}\mathcal{R}}^\alpha}{c_{\mathcal{S}\mathcal{R}}S_\mathcal{S}}$ and $Y_2\triangleq\frac{a^{D-\delta-d_{\mathcal{S}\mathcal{R}}}(D-\delta-d_{\mathcal{S}\mathcal{R}})^\alpha}{c_{\mathcal{R}\mathcal{D}}S_\mathcal{R}}$. So,~\eqref{L_var} and~\eqref{Bh_var} along with the implicit negativity of $2\times2$ leading principal submatrix of $B_H(\overline{\gamma})$ complete the proof.
    \color{black} 
     
\bibliographystyle{IEEEtran}
\bibliography{references_COMML}

\end{document}